\documentstyle[12pt]{article}
\title{ $N\bar N, \bar\Delta N, \Delta \bar N$ excitation for 
          pion propagator \\ 
          in nuclear matter}
\author{Liang-gang Liu$^{1,2}$, 
Xiang-qian Luo$^{1,2}$,Qi-fa Zhou$^2$, Wei Chen$^2$  \\
$^1$CCAST (World Laboratory), P. O. Box 8730, Beijing 100080 \\
$^2$ Department of Physics, Zhongshan University \\
 Guangzhou, Guangdong 510275, P. R. China \\ 
Masahiro Nakano \\
University of Occupational and Environmental Health\\ 
  Kitakyusyu 807, Japan}
\date{}
\begin{document}
\maketitle
\vspace{2cm}

\begin{abstract}
 The  particle - hole and delta - hole excitations are well known elementary 
excitation modes for pion propagator in nuclear matter. But, the excitation
also involves antiparticles, namely, nucleon-antinucleon, antidelta-nucleon 
and delta-antinucleon excitations. These are important for high 
energy-momentum as well,
and not studied before. In this paper, we give both the formulae and the 
numerical calculations for the real and the imaginary parts of these 
excitations.
\end{abstract}
\par
PACS number(s): 21.65.+f, 13.75.Gx, 11.10.Gh

\newpage
The excitations caused by the meson propagation in nuclear matter, or the
polarization insertion of the meson propagators, is very elementary and
important, related to many fields in nuclear physics [1, 2], e. g.,
the meson propagators are indispensable to the calculation of 
the binding energy of nuclear matter in the RPA approximation and in  
relativistic 
many-body field theory [4]. Taking the pion as another example, 
its propagator 
has a decisive effect on the dilepton production rate in heavy ion collisions 
[3]. But in the calculation of the dilepton production rate, the pion 
propagator used has been nonrelativistic. The validity of using a
nonrelativistic pion propagator for the rho meson production 
energy-momentum region is 
obviously suspicious. A relativistic description of the particle-hole ({\it
ph}) excitation for the meson propagator was not available until our last work 
[5, 6]. In that work, we pointed out that the {\it ph} and 
particle-antiparticle ($N\bar N$) contributions could not be split by using 
the traditional method [7, 8]. We put forth a new method and calculated the
dimensonic function [5] and pion propagator [6] with an analytical expression
for relativistic {\it ph} excitation. We have extended this formalism to study
the pion propagator in nuclear matter [9]. Besides {\it ph} excitation, the 
delta-hole ($\Delta h$) excitation channel will open and play a dominant 
role for pion propagator. In all these calculations, traditional or new 
formalism, the $N\bar N$ excitation antidelta-nucleon ($\bar \Delta N$),
 and delta-antinucleon ($\Delta \bar N$) excitations were ignored, and nobody
has ever studied them.

In this paper, we will study these excitations with antiparticles involved, 
by calculating the imaginary part of the polarization insertion of pion 
propagator. The nucleon and $\Delta$-isobar propagators in nuclear matter
$S(p), S_{\mu\nu}(p)$ can be expressed  in terms of a particle
-antiparticle formalism as follows:
\begin{eqnarray}
S(p) = S_{p}(p) + S_{h}(p) + S_{\bar p}(p)
\end{eqnarray}
\begin{eqnarray}
S_{p}(p) = (1 - n_{\bf p})
 \frac{1}{2E_N({\bf p})} \frac{\gamma_{0}E_N({\bf p}) - \vec{\gamma}
\cdot {\bf p} + {\tilde m}_N}{p_{0} - E_N({\bf p}) + i \epsilon},
\end{eqnarray}
\begin{eqnarray}
S_{h}(p) = \frac{n_{\bf p}}{2E_N({\bf p})} \frac{\gamma_{0}
E_N({\bf p}) - \vec{\gamma} \cdot {\bf p} + 
{\tilde m}_N}{p_{0} - E_N({\bf p}) - i \epsilon},
\end{eqnarray}
\begin{eqnarray}
S_{\bar p}(p) =  - \frac{1}{2E_N({\bf p})} \frac{-\gamma_{0}E_N({\bf p}) + 
\vec{\gamma} \cdot {\bf p} + {\tilde m}_N}{p_{0} + 
E_N({\bf p}) - i \epsilon},
\end{eqnarray}
where $n_{\bf p}$ is the nucleon distribution function and 
$S_p, S_h, S_{\bar p}$ are the particle, hole, antiparticle propagators. Also,
\begin{eqnarray}
S^{\mu \nu}(p) = S^{\mu \nu}_{\Delta}(p) + S^{\mu \nu}_{\bar \Delta}(p),
\end{eqnarray}
\begin{eqnarray}
S^{\mu \nu}_{\Delta}(p) = \frac{{\tilde m}_\Delta}{E_\Delta({\bf p})}
\cdot \frac{\Lambda^{\mu \nu}_{+}({\bf p})}{p_{0} - E_\Delta({\bf p}) 
+ i\epsilon},
\end{eqnarray}
\begin{eqnarray}
S^{\mu \nu}_{\bar \Delta}(p) = - \frac{{\tilde m}_\Delta}{E_\Delta({\bf p})}
\cdot \frac{\Lambda^{\mu \nu}_{-}( - {\bf p})}{p_{0} + E_\Delta({\bf p}) 
- i\epsilon},
\end{eqnarray}
where  $S^{\mu \nu}_{\Delta}(p),  S^{\mu \nu}_{\bar \Delta}(p)$ are the delta
and antidelta isobar propagator, respectively, and 
$\Lambda^{\mu \nu}_{\pm}({\bf p})$ is the corresponding projection operator:
\begin{eqnarray}
\Lambda^{\mu \nu}_{+}({\bf p}) = - \frac{\gamma \cdot p + {\tilde m}_{\Delta}}
{2 {\tilde m}_{\Delta}} \cdot [g^{\mu \nu} - \frac{1}{3} \gamma^{\mu} 
\gamma^{\nu} - \frac{2}{3 {\tilde m}^{2}_{\Delta}}p^{\mu}p^{\nu} + \frac{1}
{3 {\tilde m}_{\Delta}}(p^{\mu}\gamma^{\nu} - p^{\nu}\gamma^{\mu})],
\end{eqnarray}
\begin{eqnarray}
\Lambda^{\mu \nu}_{-}({\bf p}) = - \frac{- \gamma \cdot p + {\tilde m}
_{\Delta}}
{2 {\tilde m}_{\Delta}} \cdot [g^{\mu \nu} - \frac{1}{3} \gamma^{\mu} 
\gamma^{\nu} - \frac{2}{3 {\tilde m}^{2}_{\Delta}}p^{\mu}p^{\nu} - \frac{1}
{3 {\tilde m}_{\Delta}}(p^{\mu}\gamma^{\nu} - p^{\nu}\gamma^{\mu})].
\end{eqnarray}
Here and in eqs. (2-4), $\tilde m_N$,
${\tilde m}_{\Delta}$ is effective mass of nucleon and $\Delta$-isobar 
in nuclear matter, and  $E_{N({\Delta})} ({\bf p}) = \sqrt{{\bf p}^{2} + 
{\tilde m}^{2}_{N(\Delta)}}$. 

The polarization with the nucleon and nucleon-delta loop insertions 
for the pion propagator can be calculated in the 
standard field theory, and it reads:
\begin{eqnarray}
\Pi(q)  = \Pi_N(q) + \Pi_\Delta(q),
\end{eqnarray}
\begin{eqnarray}
\Pi_N(q) = - 2i \frac{f_{\pi NN}^{2}}{m_\pi^2} q_\mu q_\nu 
\int \frac{d p}{(2\pi)^4} Tr [\gamma_5 \gamma^\mu S(p) 
\gamma_5 \gamma^\nu S(p + q)],
\end{eqnarray}
\begin{eqnarray}
\Pi_\Delta (q) = - {8i \over 6} \frac{f_{\pi NN}^{2}}{m_\pi^2} q_\mu q_\nu 
\int \frac{d p}{(2\pi)^4} Tr [S(p) S^{\mu \nu}(p + q)] + (q \rightarrow - q),
\end{eqnarray}
where $\Pi_N(q), \Pi_\Delta(q)$ are due to nucleon and nucleon-delta loop
contributions, respectively. We have used the pseudo-vector coupling for
$\pi NN$ and $\pi N \Delta$ interacting vertices, 
$f_{\pi NN}$ and $f_{\pi N \Delta}$ 
are the corresponding coupling constants. $f_{\pi N \Delta}$ = 
2 $f_{\pi NN}$ with $f_{\pi NN}$ = 0.988 [8]. In ref. [9], we have explained
why the arbitrary parameter $\xi$ that appeared in ref. [8] does not appear in our 
formalism.

Substitute eqs. (2-4, 5-9) into eqs. (11, 12), and the polarization insertion
can be expressed in terms of nothing but the {\it ph}, $N\bar N$, $\Delta h$,
$\bar \Delta N$ and $\Delta \bar N$ excitation contributions:
\begin{eqnarray}
\Pi_N(q) = \Pi_{ph}(q) + \Pi_{N\bar N}(q),
\end{eqnarray}
\begin{eqnarray}
\Pi_\Delta (q) = \Pi_{\Delta h}(q) + \Pi_{\bar\Delta N}(q) + 
\Pi_{\Delta \bar N}(q), 
\end{eqnarray}
the subscript denoting the corresponding excitation. The {\it ph} and $\Delta
h$ excitations have been studied extensively in previous work [6, 9], so we
will pay more attention to antiparticle excitation in this paper. The 
imaginary part of these polarization insertions can be written as follows:
\begin{eqnarray}
Im \Pi_{ph} (q) & = & 4 \frac{f_{\pi NN}^{2}}{m_\pi^2} {\tilde m}_N^2 q^2 \pi
\int {d{\bf p} \over (2\pi)^3
E_N({\bf p}) E_N({\bf p + q})} (1 - n_{\bf p}) n_{\bf p + q} \\ \nonumber
&  & \cdot [ \delta(E_N({\bf p}) - E_N({\bf p + q}) - q_0) 
+ \delta(E_N({\bf p}) - E_N({\bf p + q}) + q_0) ],
\end{eqnarray}
\begin{eqnarray}
Im \Pi_{N\bar N} (q) & = & - 4 \frac{f_{\pi NN}^{2}}{m_\pi^2} {\tilde m}_N^2 q^2 \pi
\int {d{\bf p} \over (2\pi)^3
E_N({\bf p}) E_N({\bf p + q})} (1 - n_{\bf p})  \\ \nonumber
&  & \cdot [ \delta(E_N({\bf p}) + E_N({\bf p + q}) - q_0) 
+ \delta(E_N({\bf p}) + E_N({\bf p + q}) + q_0) ],
\end{eqnarray}
\begin{eqnarray}
Im \Pi_{\Delta h} (q) & = &  {4\over 9} \frac{f_{\pi NN}^{2}}{m_\pi^2} 
[({\tilde m}_N + {\tilde m}_\Delta)^2 - q^2] \cdot [q^2 - {1\over 4{\tilde m}
^2_\Delta} ({\tilde m}_\Delta^2 - {\tilde m}_N^2 + q^2)^2] \cdot \\ \nonumber
& & \pi \int {d{\bf p} \over (2\pi)^3
E_\Delta({\bf p}) E_N({\bf p + q})} n_{\bf p + q} \cdot
[ \delta(E_\Delta({\bf p}) - E_N({\bf p + q}) - q_0) \\ \nonumber 
& & + \delta(E_\Delta({\bf p}) - E_N({\bf p + q}) + q_0) ],
\end{eqnarray}
\begin{eqnarray}
Im \Pi_{\bar \Delta N} (q) & = & - {4\over 9} \frac{f_{\pi NN}^{2}}{m_\pi^2} 
[({\tilde m}_N + {\tilde m}_\Delta)^2 - q^2] \cdot [q^2 - {1\over 4{\tilde m}
^2_\Delta} ({\tilde m}_\Delta^2 - {\tilde m}_N^2 + q^2)^2] \cdot \\ \nonumber
& & \pi \int {d{\bf p} \over (2\pi)^3
E_\Delta({\bf p}) E_N({\bf p + q})}(1 - n_{\bf p + q}) \cdot
[ \delta(E_\Delta({\bf p}) + E_N({\bf p + q}) - q_0)  \\ \nonumber 
& &+ \delta(E_\Delta({\bf p}) + E_N({\bf p + q}) + q_0) ],
\end{eqnarray}
\begin{eqnarray}
Im \Pi_{\Delta \bar N} (q) & = & - {4\over 9} \frac{f_{\pi NN}^{2}}{m_\pi^2} 
[({\tilde m}_N + {\tilde m}_\Delta)^2 - q^2] \cdot [q^2 - {1\over 4{\tilde m}
^2_\Delta} ({\tilde m}_\Delta^2 - {\tilde m}_N^2 + q^2)^2] \cdot \\ \nonumber
& & \pi \int {d{\bf p} \over (2\pi)^3
E_\Delta({\bf p}) E_N({\bf p + q})} \cdot
[ \delta(E_\Delta({\bf p}) + E_N({\bf p + q}) - q_0)  \\ \nonumber 
& & + \delta(E_\Delta({\bf p}) + E_N({\bf p + q}) + q_0) ].
\end{eqnarray}

The nucleon distribution function and its combination correctly indicates the
excitation properties of each excitation above. But this is not the case for
traditional approach[6, 8]. The variables in the $\delta$ functions in the 
integrands gives the thresholds and the boundary of the energy-momentum for
the corresponding excitations. In Fig. 1, we show the imaginary part as a 
\begin{center}
\begin{tabular}{|c|} \hline
Figure 1 \\ \hline
\end{tabular}
\end{center}
function of energy $\omega$ for fixed momentum $q = 2.5 k_F$, where $k_F =
1.42 fm^{-1}$. The effective mass of nucleon is derived from the calculation
of the binding energy of nuclear matter in the relativistic Hartree 
approximation [10]. That is $x \equiv {{\tilde m}_N \over m_N}$ = 0.72 ($m_N$
is nucleon mass in free space). We set $x_\Delta \equiv {{\tilde m}_\Delta 
\over m_\Delta}$ = 1 ($m_\Delta$ is the $\Delta$-isobar mass in free space)
for this calculation. The thresholds and the magnitudes for antiparticle
excitation are  larger than {\it ph} and $\Delta h$ excitations. For 
{\it ph} and $\Delta h$ excitations, they are nonvanishing only in a very 
limited  region of energy. But for antiparticle excitations, once the energy
is larger than the thresholds,  no matter how large it is, the corresponding
excitation will not vanish. The $\Delta \bar N$ and $\bar \Delta N$ 
excitations, which start from the same threshold, are hard to distinguish from 
each other. We also find that all these excitations depend on the effective
mass of nucleon and delta very strongly. In ref. [11], we calculated the binding 
energy with pion and $\Delta$-isobar degree of freedom explicitly included,
and we found $x = 0.905, x_\Delta = 0.928$ at $k_F = 1.42 fm^{-1}$. The imaginary
parts are calculated again by use of these parameters and shown in Fig. 2. 
\begin{center}
\begin{tabular}{|c|} \hline
Figure 2 \\ \hline
\end{tabular}
\end{center}
Comparing with Fig. 1, we can see that the $\Delta h$ excitation moves to the 
left but those of antiparticles move to the right apparently. The strength
of the {\it ph} and $\Delta h$ are also stronger than those in Fig. 1.

The real and the imaginary part are related to each other by the dispersion 
relation:
\begin{eqnarray}
Re \Pi (q_0, {\bf q})  =  
\frac{P}{\pi} \int_0^\infty  d\omega^2 
{Im \Pi (\omega^2, {\bf q}) \over \omega^2 - q_0^2}.
\end{eqnarray}
For antiparticle excitations, $Re \Pi$ is divergent due to the fact that 
$Im \Pi \neq 0$ and is proportional to $\omega^2$ at least
for $\omega \rightarrow \infty$; it is also nonrenormalizable 
for pseudo-vector coupling in the sense
of the standard counter-term renormalization scheme. On the other hand, however,
nucleon and $\Delta$-isobar are not bare particles, their  
structure are strongly
energy-momentum dependent instead. This dependence is described by $\pi NN$
and $\pi \Delta N$ vertex functions, or form factors (see ref. [12] for 
example). For the space-like region, the form factor can be described by 
monopoles or dipoles which are widely used for baryon-baryon interactions
[13, 14]. But in the time-like region, there are some problems with these 
multi-pole form factors especially for large energy-momentum transfer [15].
The exponential form factor is also a good candidate[4, 16]; it 
resembles multi-pole for low energy-momentum but converges more quicker for
high energies. It has the following form:
\begin{eqnarray}
F(q^2) = e^{ - {q^2 \over \Lambda^2}} \theta (q^2) + 
e^{  {q^2 \over \Lambda^2}} \theta (- q^2),
\end{eqnarray}
where $\Lambda$ is the cut-off parameter. Generally $\Lambda$
it is about $7 m_\pi$
[4, 17], but this was a too low value $2.7 m_\pi$ in ref. [16]. We choose
$\Lambda = 6.8m_\pi$ in this paper. Substituting $f_{\pi NN}$
with $f_{\pi NN} F(q^2)$ in eqs. (16), $f_{\pi N\Delta}$ with 
$f_{\pi N\Delta} F(q^2)$ in eqs. (18, 19) respectively, the corresponding
real parts calculated by eq. (20) are shown in Fig. 3. The other parameters
\begin{center}
\begin{tabular}{|c|} \hline
Figure 3 \\ \hline
\end{tabular}
\end{center}
are the same as in Fig. 1. We can see that the magnitude of $Re \Pi$ for
$N\bar N$ excitation is about $10^2$ larger than $\bar \Delta N$ and 
$\Delta \bar N$ excitations. This is due to the energy threshold for
$N\bar N$ excitation being smaller than the latter as shown in Fig. 1, 
resulting in large suppression by the form factor in the large energy region. 
We also find that instead of the exponential form factor (21), the monopole
form factor [12, 13, 14] does not converge quickly enough to make eq. (20) finite,
except for a very small cut-off parameter. This is the same case as mentioned 
in ref. [16]. That means that measurements and theoretical studies for the 
nucleon strong interaction form factor in the time-like region and in the
larger energy transfer region are very necessary indeed, since it plays a key 
role in obtaining the real part of the polarization insertions.

In this paper, we neglect the natural width of the $\Delta$-isobar 
propagator, but this will be remedied in future work.
Another related issue on the renormalization of the real part of the 
polarization insertion is the renormalization scheme. For nonrenormalizable
pseudo-vector coupling, our scheme adopted in this paper is to use the 
dispersion relation with the introduction of a form factor with a cut-off
parameter $\Lambda$. An alternative scheme is to introduce a cut-off
momentum in momentum space $\Lambda_p$ [18]. The renormalized polarization 
insertion thus obtained depends on $\Lambda_p$ also, so it will be very 
interesting to compare two schemes and study the difference. The work along 
line is in progress.

This work supported in part by the National Natural Science Foundation of
China.

\newpage
\section*{Reference}
\begin{description}
\item
[[1]] E. Oset, H. Toki {\scriptsize AND} W. Weise, Phys. Rep. 
{\bf 83}, 281 (1982).
\item
[[2]]G. Chanfray, {\it Pion in the nuclear medium}, Preprint LYCEN 9623, 1996.
\item
[[3]]J. Helgesson {\scriptsize AND} J. Randrup, Phys. Rev. 
       {\bf C52}, 427 (1995).
\item
[[4]]H. Jong, G. A. Miller, Phys. Rev. {\bf C43}, 1958 (1991).
\item
[[5]]L. G. Liu, Q. F. Zhou {\scriptsize AND} T. S. Lai, Phys. Rev. {\bf C51},
    R2302 (1995).
\item
[[6]]L. G. Liu, Phys. Rev. {\bf C51}, 3421 (1995).
\item
[[7]]V. F. Dmitriev {\scriptsize AND} Toru Suzuki, Nucl. Phys. {\bf A438},
     697 (1985).
\item
[[8]]T. Herbert, K. Wehrberger, {\scriptsize AND} F. Beck, Nucl. Phys. 
   {\bf A 541}, 699 (1993).
\item
[[9]]L. G. Liu, {\scriptsize AND} M. Nakano, {\it The dispersion relation
     relation of pion in nuclear matter}, to appear in Nucl. 
     Phys. {\bf A} (June, 1997).
\item
[[10]]S. A. Chin, Ann. Phys. {\bf 108}, 301 (1977).
\item
[[11]]L. G. Liu, Q. F. Zhou, Z. Phys. {\bf A357}, 27 (1997).
\item
[[12]]T. W. Durso, A. D. Jackson, B. J. Verwest, Nucl. Phys. {\bf A282},
      404 (1977).
\item
[[13]]K. Holinde, Phys. Rep. {\bf 68}, 121 (1981).
\item
[[14]]J. Flender, M. F. Gari, Phys. Rev. {\bf C51}, R1619 (1995).
\item         
[[15]]K. Bongardt, H. Pilkuhn, H. G. Schlaile, Phys. Lett. {\bf B52},
      271 (1974).
\item
[[16]]C. L. Korpa, R. Malfliet, Phys. Rev. {\bf C52}, 2756 (1995).
\item
[[17]] L. H. Xia, C. M. Ko, L. Xiong {\scriptsize AND} J. Q. Wu, 
   Nucl. Phys. {\bf A485}, 721 (1988).
\item
[[18]]M. Nakano et al,  Phys. Rev. {\bf C55}, 1 (1997). 
\end{description}

\newpage
\section*{Figure Captions}
\begin{description}
\item
[Figure 1.]
$- Im \Pi (\omega, q)$ (in unit of $m_\pi^2$) versus energy $\omega$
(in unit of $m_\pi$) for fixed momentum $q = 2.5 k_F$. The meaning of 
the letters which are explained in the text,
indicate the corresponding excitations.
\item
[Figure 2.]
The same as Fig. 1 but with effective mass of nucleon and delta different
as described in the text.
\item
[Figure. 3.]
The real part of the polarization insertion 
- $Re \Pi$ (in unit of $m^2_\pi$) versus energy $\omega$ (in unit of $m_\pi$)
for fixed momentum $q = 2.5 k_F$. The left axis is for $N\bar N$ (solid line),
the right axis is for $\bar \Delta N$ (dotted line) and $\Delta \bar N$
(dashed line) excitations as indicated by the letters. The right axis is
multiplied by factor $10^3$.
\end{description}
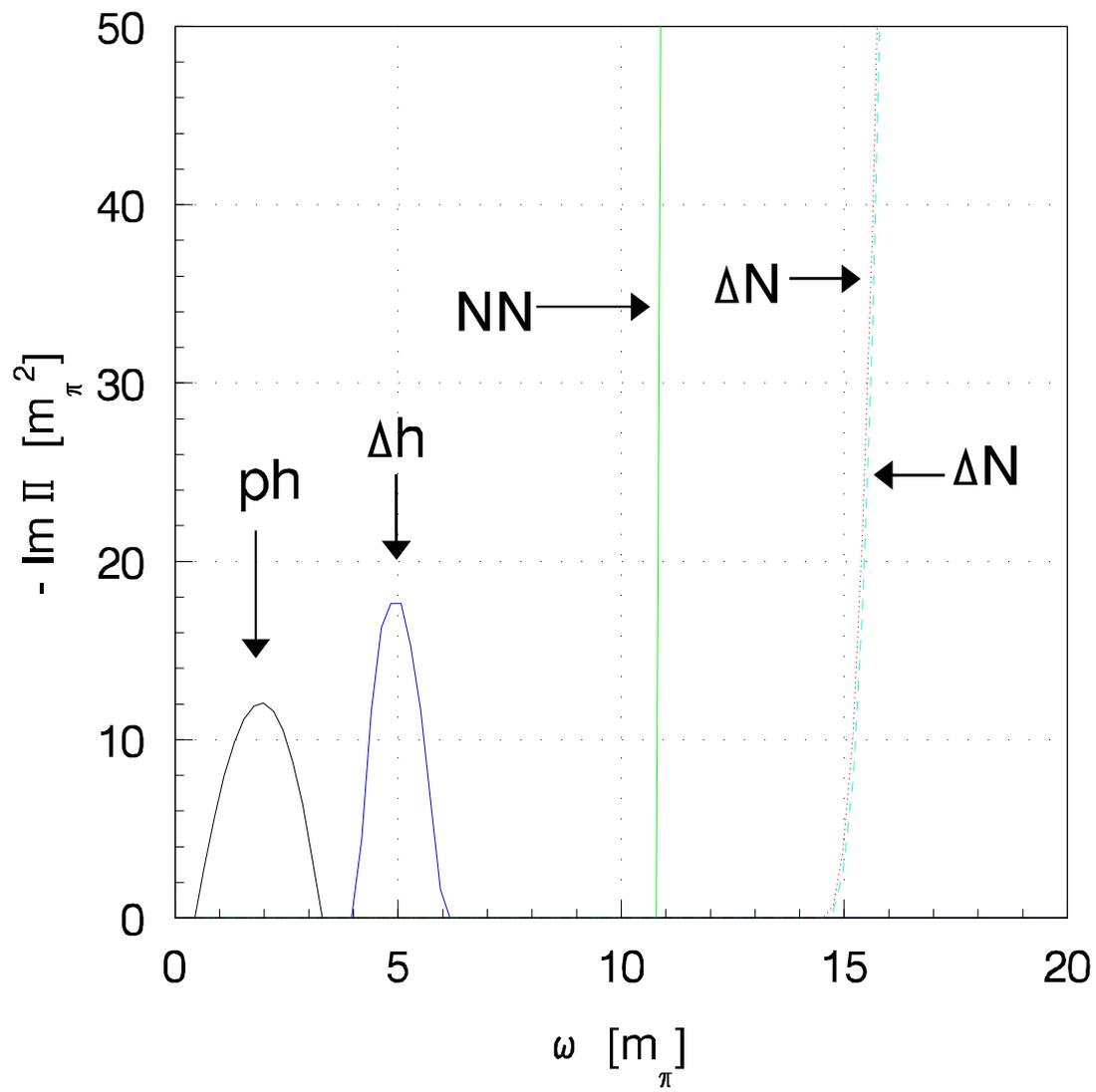
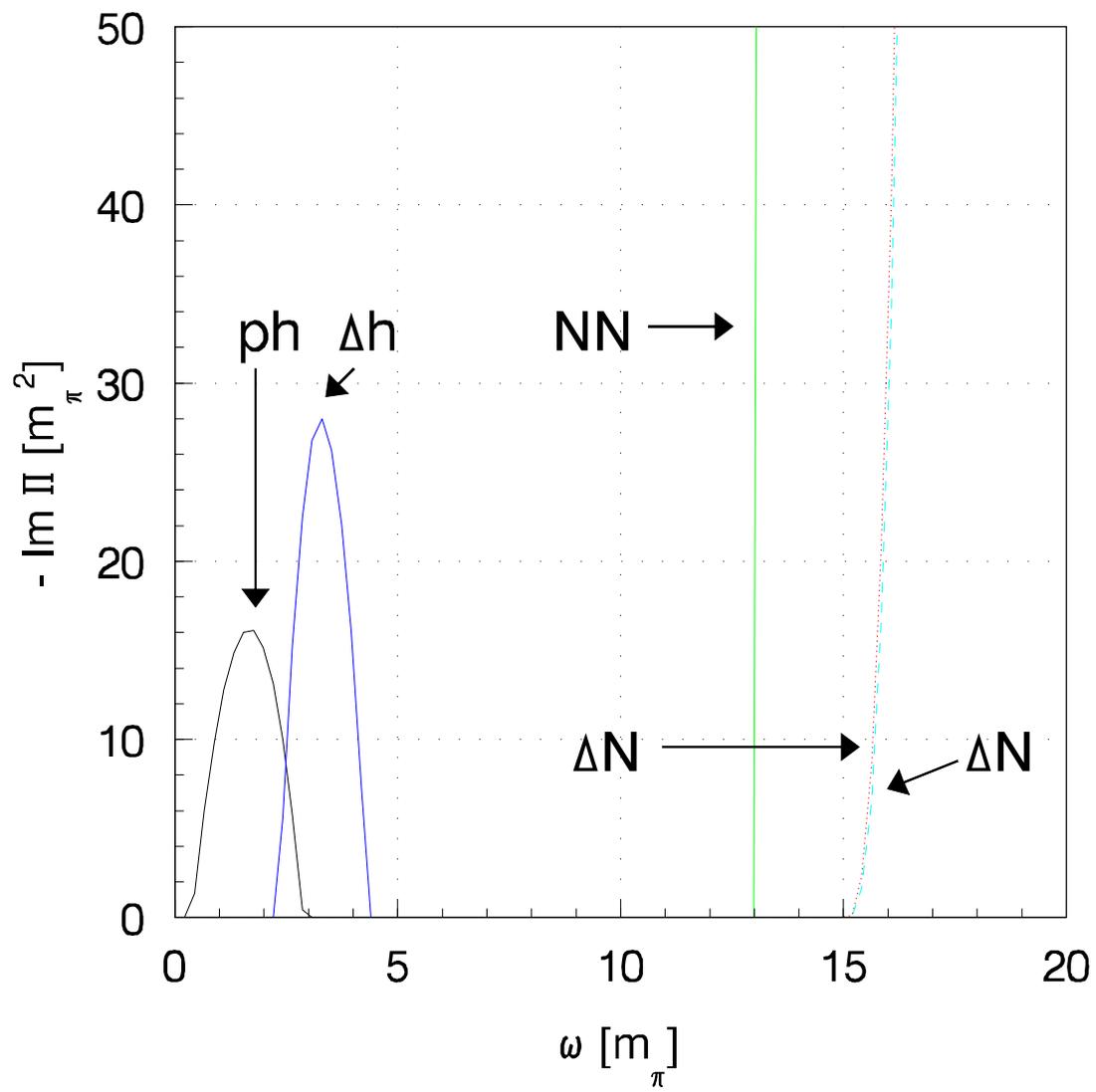
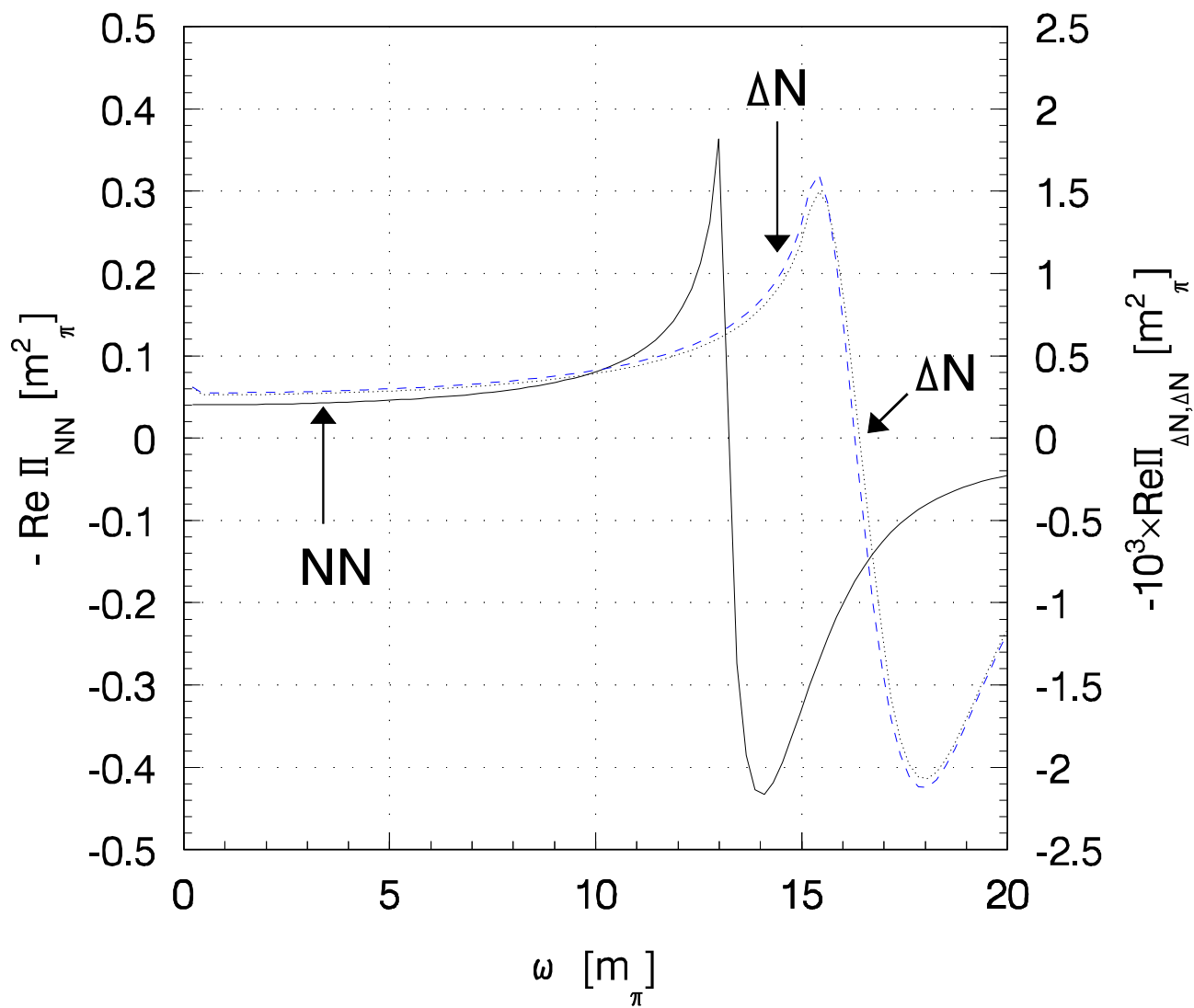
\end{document}